\documentclass[%
 reprint,
superscriptaddress,
 amsmath,amssymb,
 aps,
]{revtex4-1}

\usepackage{graphicx}
\usepackage{dcolumn}
\usepackage{bm}

\begin{document}

\preprint{APS/123-QED}

\title{Precise Frequency Measurement of the 2$^{1}$S$_{0}$-3$^{1}$D$_{2}$ Two-Photon Transition in atomic $^{4}$He}

\author{Yi-Jan Huang}
\affiliation{
 Institute of Photonics Technologies, National Tsing Hua University, Hsinchu 30013, Taiwan
}
\author{Yu-Chan Guan}
\affiliation{
 Institute of Photonics Technologies, National Tsing Hua University, Hsinchu 30013, Taiwan
}
\author{Yao-Chin Huang}
\affiliation{
Department of Physics, National Tsing Hua University, Hsinchu 30013, Taiwan
}
\author{Te-Hwei Suen}
\affiliation{
Department of Physics, National Tsing Hua University, Hsinchu 30013, Taiwan
}
\author{Jin-Long Peng}
\affiliation{
Center for Measurement Standards, Industrial Technology Research Institute, Hsinchu 30011, Taiwan
}
\author{Li-Bang Wang} \email{lbwang@phys.nthu.edu.tw}
\affiliation{
Department of Physics, National Tsing Hua University, Hsinchu 30013, Taiwan
}

\author{Jow-Tsong Shy}
 \email{shy@phys.nthu.edu.tw}
\affiliation{
 Institute of Photonics Technologies, National Tsing Hua University, Hsinchu 30013, Taiwan
}

\affiliation{
Department of Physics, National Tsing Hua University, Hsinchu 30013, Taiwan
}

\date{\today}

\begin{abstract}
We present the first precise frequency measurement of the 2$^{1}$S$_{0}$-3$^{1}$D$_{2}$ two-photon transition in $^{4}$He at 1009~nm. The laser source at 1009~nm is stabilized on an optical frequency comb to perform the absolute frequency measurement. The absolute frequency of 2$^{1}$S$_{0}$-3$^{1}$D$_{2}$ transition is experimentally determined to be 594~414~291~803(13)~kHz with a relative uncertainty of 1.6 $\times$ 10$^{-11}$ which is more precise than previous determination by a factor of 25. Combined with the theoretical ionization energy of the 3$^{1}$D$_{2}$ state, the ionization energy of the2$^{1}$S$_{0}$  state is determined to be 960~332~040~866(24) kHz. In addition, the deduced 2$^{1}$S$_{0}$ and 2$^{3}$S$_{1}$  Lamb shifts are 2806.817(24) and 4058.8(24) MHz respectively which are 1.6 times better than previous determinations.
\begin{description}
\item[Usage]
Secondary publications and information retrieval purposes.
\item[PACS numbers]
May be entered using the \verb+\pacs{#1}+ command.
\item[Structure]
You may use the \texttt{description} environment to structure your abstract;
use the optional argument of the \verb+\item+ command to give the category of each item. 
\end{description}
\end{abstract}

\pacs{Valid PACS appear here}
\maketitle


\section{\label{sec:level1}INTRODUCTION}
Helium is the simplest multi-electron atom and it plays a crucial role in testing many-body quantum electrodynamics (QED) calculations. Recently, the electronic structure of helium has been calculated to a very high precision~\cite{drake2006,drake2008,yerokhin2010,zhang2015,pachucki2017}. In particular, the metastable 2$^{1}$S$_{0}$ and 2$^{3}$S$_{1}$states of are of great experimental interest in spectroscopy studies due to their larger effects of Lamb shift and accessible light sources. To determine the Lamb shift of the 2S states, measurements of the 2S-nD two-photon transitions are typically chosen since the theoretical uncertainties of the D states are much smaller than the low-lying S and P states. For example, the Lamb shift of the 2$^{3}$S$_{1}$ state has been determined using the 2$^{3}$S$_{1}$-3$^{3}$D$_{1}$ two-photon transition~\cite{dorrer1997} and the Lamb shift of the 2$^{1}$S$_{0}$ state has been determined using the 2$^{1}$S$_{0}$ - n$^{1}$D$_{2}$  two-photon transitions (7 $<$ n $<$ 20)~\cite{lichten1991}. In addition, the high-precise frequency measurement of the 2$^{1}$S$_{0}$ - 2$^{3}$S$_{1}$~\cite{van2011} doubly forbidden transition at 1557 nm serves as a bridge to connect the triplet and singlet states. For example, the Lamb shift of the 2$^{1}$S$_{0}$ state can be determined from the 2$^{3}$S$_{1}$-3$^{3}$D$_{1}$ transition~\cite{dorrer1997} and  2$^{3}$S$_{1}$-2$^{1}$S$_{0}$ transition~\cite{van2011}. \par
Due to the recent developments of the optical frequency comb~(OFC), several OFC-based measurements are performed to directly determine the absolute transition frequencies of He~\cite{van2011,pastor2004,pastor2012,luo2013,luo2013a,notermans2014,luo2016}. The best previous determination of the Lamb shift of the  2$^{1}$S$_{0}$ state in $^{4}$He, obtained from the $^{4}$He  2$^{3}$S$_{1}$-2$^{3}$P$_{0}$ [10],  2$^{3}$P$_{0}$-3$^{3}$D$_{1}$~\cite{luo2016} and 2$^{3}$S$_{1}$-2$^{1}$S$_{0}$ transitions~\cite{van2011}, has an uncertainty of 35 kHz. In this work, we present our precise absolute frequency measurement of the singlet 2$^{1}$S$_{0}$-3$^{1}$D$_{2}$ two-photon transition at 1009~nm. The uncertainty reaches 13 kHz and we are able to reduce the uncertainty of the ionization energy of 2$^{1}$S$_{0}$ state and the Lamb shift of both the 2$^{1}$S$_{0}$ and 2$^{3}$S$_{1}$ states to 24~kHz, which is limited mostly by the uncertainty of the theoretical ionization energy of the 3D states (20~kHz).

\section{\label{sec:level2}EXPERIMENTAL SETUP}
\begin{figure}
\includegraphics[width=0.5\textwidth]{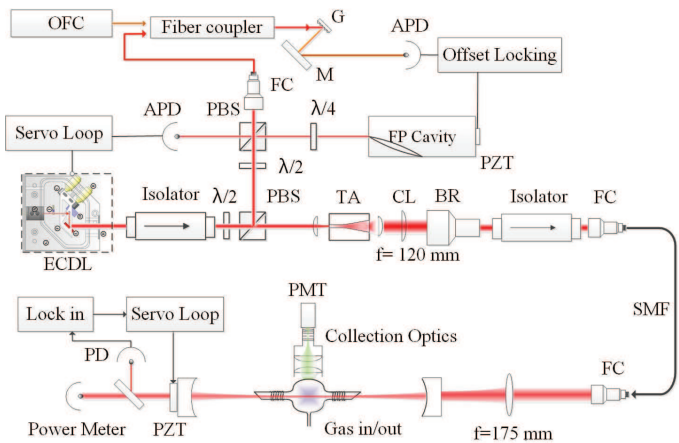}
\caption{\label{fig:setup}The schematic diagram of the experimental setup for absolute frequency measurement of the 2$^{1}$S$_{0}$-3$^{1}$D$_{2}$ two-photon transition in $^{4}$He. The output of the 1009~nm external-cavity diode laser (ECDL) is divided into three beams. One is referenced to a fiber-based optical frequency comb (OFC) via a fiber link for absolute frequency measurement, one is used to pre-stabilize the ECDL frequency with a Fabry-P$\acute{e}$rot (FP) cavity, and the remaining beam is injected into the tapered amplifier (TA) to boost the laser power. We employ cavity enhancement absorption for the two-photon spectroscopy with the He gas cell inside a cavity. The PMT monitors the fluorescence signal from 3$^{1}$D$_{2}$ to 2$^{1}$P$_{1}$ at 668~nm through the collection optics. }
\end{figure}
The schematic of our experimental setup is shown in Fig.~\ref{fig:setup}. The 1009~nm light source used for the 2$^{1}$S$_{0}$-3$^{1}$D$_{2}$ two-photon transition in $^{4}$He is a homemade external-cavity diode laser (ECDL). The laser is pre-stabilized onto a Fabry-P$\acute{e}$rot (FP) cavity by the Pound-Drever-Hall locking technique~\cite{drever1983}, and its laser linewidth is reduced to 250~kHz in 1~ms intergration time observed using the beat note between ECDL and one comb line of an OFC. The OFC is based on a femtosecond Er-doped fiber laser~\cite{peng2007}.  For our experiment, we amplify its output and extend its supercontinuum spectrum from 1050~nm to 980~nm by a high nonlinear fiber. Its repetition rate (f$_{r}$, 250 MHz) and offset frequency (f$_{o}$) are both locked to frequency synthesizers referenced to a cesium clock (Microsemi. 5071A). The accuracy of the OFC is better than 1~$\times$ 10$^{-12}$ at 1000~sec. integration time. The laser frequency is locked on one comb line of the OFC by offset locking technique. To increase the optical power for the two-photon transition, the output power of the ECDL is boosted to 2~W by a tapered amplifier (TA). After passing through an optical isolator and a single-mode optical fiber (SMF), we have a 750~mW laser beam with TEM$_{00}$ profile for the experiment. \par
 Helium gas is filled in a glass cell and a radio-frequency (RF) discharge populates the 2$^{1}$S$_{0}$ metastable state. The glass cell is a sphere (6~cm in diameter) sealed with Brewster windows at both ends. A large spherical cell is necessary to avoid quenching of the metastable atoms due to collision on the wall. The glass cell is pumped by a turbomolecular pumping system to a pressure below 10$^{-6}$~Torr and its pressure can be kept below 10$^{-6}$~Torr by a getter (SAES CapaciTorr D 400-2) over one week after the cell is sealed. When the cell is filled with $^{4}$He (typically 30-200~mTorr), the variation of the helium pressure is less than 1~$\%$ over 2 weeks with RF discharge on. To enhance the optical power to excite the two-photon transition and to obtain perfect overlapping counter-propagating laser beams, the cell is placed in a power built-up cavity formed by two spherical mirrors (200~mm radius of curvature, and 350~mm apart). One mirror is mounted on a PZT for cavity length tuning. To keep the cavity resonant with the laser frequency, the length of the cavity is locked on the transmission peak by dithering its PZT and the transmitted optical signal is demodulated using a lock-in amplifier. To eliminate any possible Zeeman effect, the Earth magnetic field is shielded by a $\mu$-metal box. The two-photon transition is detected by monitoring the 3$^{1}$D$_{2}$-2$^{1}$P$_{1}$ fluorescence at 668~nm which is collected with an optical system consisting of two f~=~6~cm lenses (3 inches in diameter), an NIR filter, a 25.4~mm aperture, a 675~nm band-pass filter (30~nm FWHM) and a f~=~50~mm lens in front of a photomultiplier tube (PMT). To avoid parasitic light of the RF discharge, the discharge is pulsed at 10~kHz and the detection is carried out in the afterglow regime. The RF power is first switched on for a 5~$\mu$s duration, and then the PMT signal is collected for a 40~$\mu$s duration at 20~$\mu$s after the RF is turned off. The laser frequency is scanned step by step with an interval of approximate 1.6~MHz by tuning the repetition rate of the OFC. The fluorescence signal and fb, beat frequency between the ECDL and the OFC,  are recorded by the computer for 6 seconds for each frequency step. The absolute frequency of the laser is given by: f = n$\times$f$_{r}$+f$_{o}$+f$_{b}$, where n is the mode number of the OFC. In the meantime, the laser frequency is monitored by a wavelength meter (HighFinesse WD30) with a resolution of 10 MHz to assist us to determine the OFC mode number~n. The working timing sequence is accurately controlled by computer.
\begin{figure}
\includegraphics[width=0.5\textwidth]{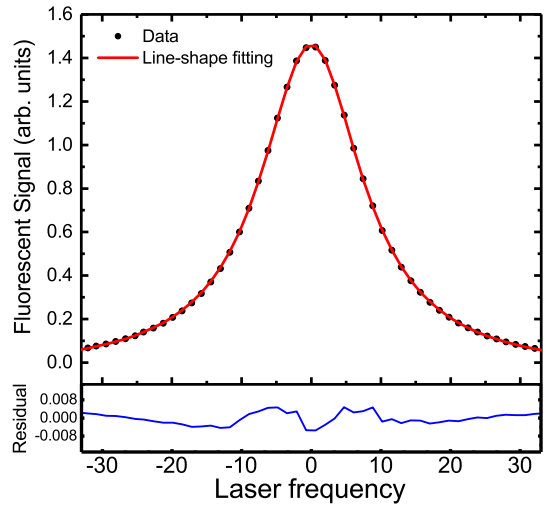}
\caption{\label{fig:spec}A typical spectrum of the 2$^{1}$S$_{0}$-3$^{1}$D$_{2}$ two-photon transition in $^{4}$He. Here, optical power inside the power built-up cavity is 15~W, the pressure in the cell is 142.5~mTorr, and the RF discharge power is 3~W. A Lorentzian line shape (red line) is used to fit the observed spectrum.}
\end{figure}

\section{\label{sec:level3}EXPERIMENTAL RESULTS}

\begin{figure}
\includegraphics[width=0.5\textwidth]{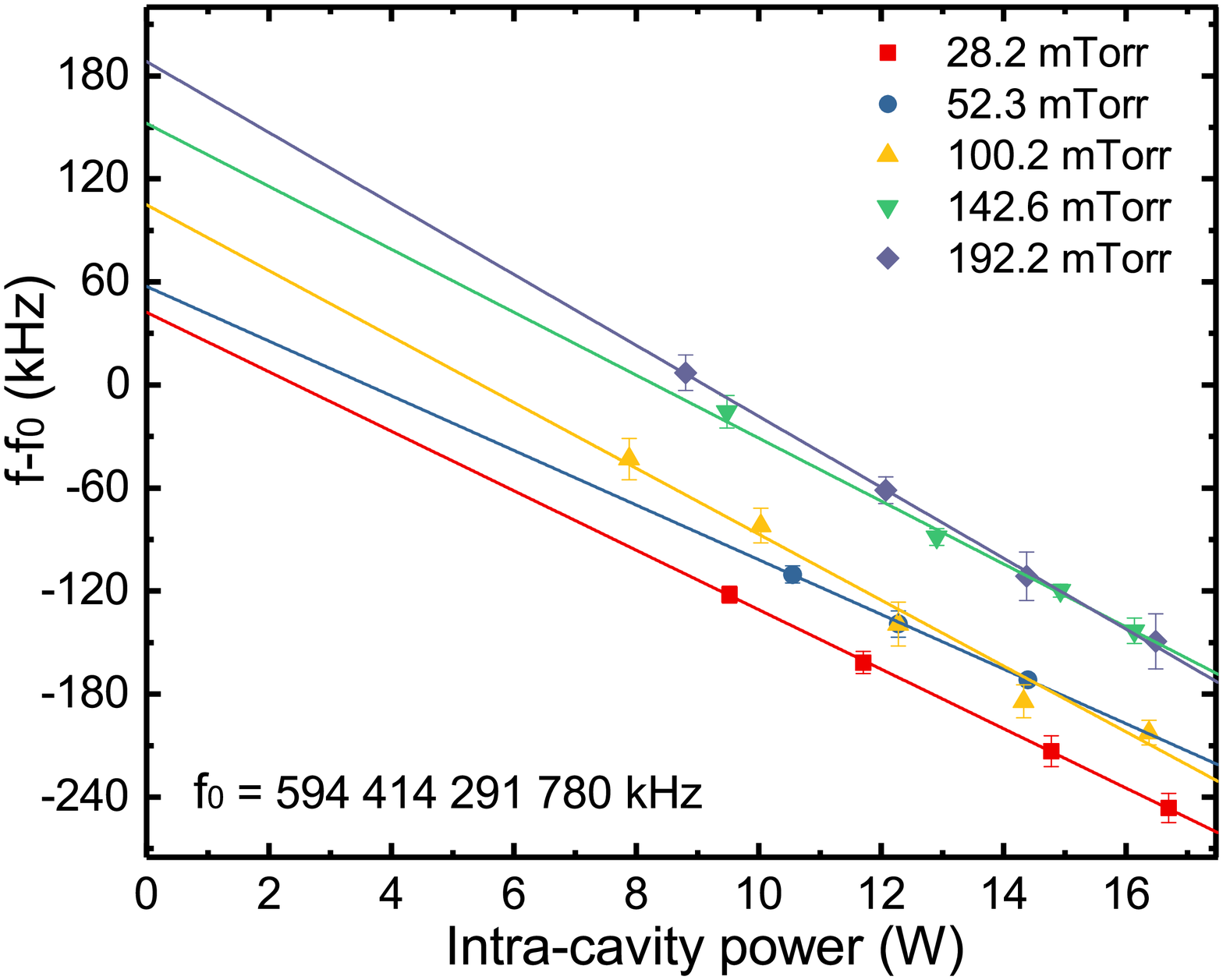}
\caption{\label{fig:fp}The extrapolations of the transition frequency versus the intra-cavity optical power at five different pressures. .}
\end{figure}
A typical spectrum of the 2$^{1}$S$_{0}$-3$^{1}$D$_{2}$ two-photon transition in $^{4}$He is shown in Fig.~\ref{fig:spec}. The locked laser frequency is scanned by tuning the repetition rate of the OFC. The spectrum has a linear background which is below  1 $\times$ 10$^{-3}$ with respect to the signal. Since the ratio between the atomic lifetime of the 3$^{1}$D$_{2}$ state and the transit time of helium atom across the laser beam at the center of the power built-up cavity is 3 approximately, according to Ref.~\cite{biraben1979}, we employ a Lorentzian line shape with a linear background to fit the spectrum, and the fitting achieves R-square $\varepsilon$~$>$~99.99 $\%$ . The fitting provides us the linewidth and the center frequency of the observed 2$^{1}$S$_{0}$-3$^{1}$D$_{2}$ transition. \par
\begin{figure}[b]
\includegraphics[width=0.5\textwidth]{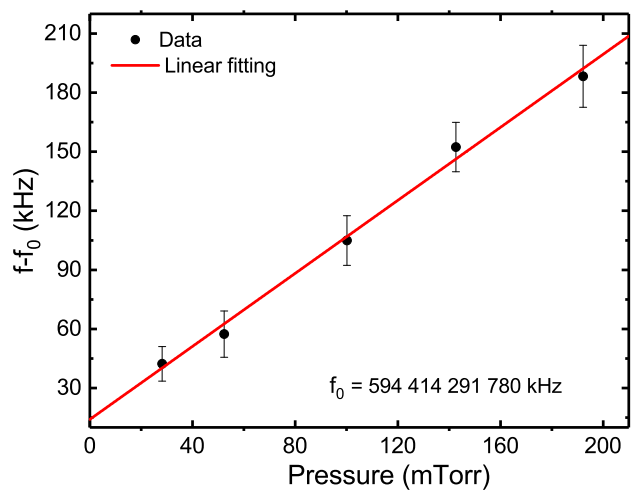}
\caption{\label{fig:fo}The extrapolations of the frequency versus the pressure. The 2$^{1}$S$_{0}$-3$^{1}$D$_{2}$ transition at zero pressure and zero power is determined to be 594~414~291~803(13) kHz.}
\end{figure}
\begin{table}[t]
\caption{\label{tab:table1}%
Uncertainty of the 2$^{1}$S$_{0}$-3$^{1}$D$_{2}$ measurement.}
\begin{ruledtabular}
\begin{tabular}{lr}
items&uncertainty\\
\hline
Statistical uncertainty&12.8~kHz\\
Second-order Doppler effect&1~kHz\\
Zeeman effect&$<$1~kHz\\
OFC accuracy (multiple a factor of 2)&4~kHz\\
Overall uncertainty&13~kHz\\
\end{tabular}
\end{ruledtabular}
\end{table}

\begin{figure}
\includegraphics[width=0.5\textwidth]{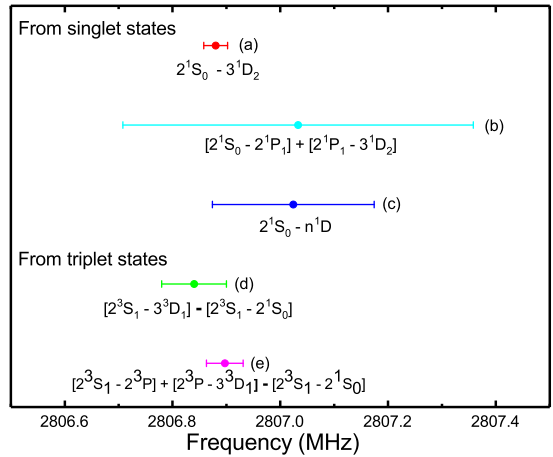}
\caption{\label{fig:comp}The comparison of experimental determination of the 2$^{1}$S$_{0}$ Lamb shift  (a) This work (b) Ref.~\cite{luo2013,luo2013a} (c) Ref.~\cite{lichten1991} (d) Ref.~\cite{dorrer1997,van2011} (e) Ref.~\cite{van2011,pastor2012,luo2016}}
\end{figure}

We have measured the spectrum for five different pressures from 28 to 192~mTorr. For each pressure, the two-photon transition frequencies were measured with intra-cavity optical power from 8 to 16~W as shown in Fig.~\ref{fig:fp}. Each data point is obtained from 12 spectrums. The absolute pressure is measured by a MKS Baratron with an accuracy better than 1$\%$ and its zero point is regularly checked at the pressure below  1 $\times$ 10$^{-6}$ Torr. Furthermore, the intra-cavity power is determined by cavity transmitting power, which is measured by a power-meter (Thorlabs 302C) with an accuracy of 3$\%$. Linear extrapolation to zero intra-cavity power allows us to correct the transition frequency shift caused by the ac-Stark shift. The ac-Stark shift has major contributions from 3D-nF states. As pressure increases, the absolute value of the slope of the transition frequency versus intra-cavity power varies from 20.6(2.2)~kHz/W to 15.9(1.5)~kHz/W. The transition frequencies at zero intra-cavity power in Fig.~\ref{fig:fp} are plotted versus pressure in Fig.~\ref{fig:fo} to determine the transition center frequency at zero power and zero pressure. The coefficient of transition frequency versus pressure is 927(92)~kHz/Torr. The effects of other experimental conditions are also investigated. The RF discharge does not have significant effect on the transition frequency and spectral linewidth as we vary the RF power from 1 to 5~W. 
\begin{figure}[b]
\includegraphics[width=0.5\textwidth]{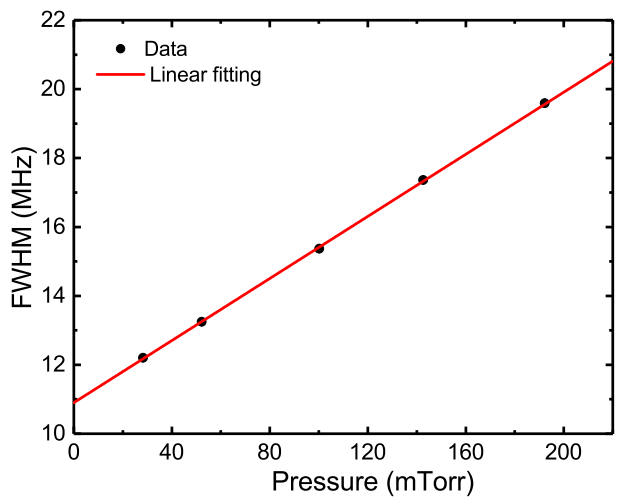}
\caption{\label{fig:fw}The extrapolations of the FWHM versus the pressure. The FWHM at zero pressure and zero power is 11.21(12)~MHz.}
\end{figure}
This can be understood since we take the data after the discharge is turned off. In addition, we measured the residual magnetic field inside the shielding $\mu$-metal box and it is below 1~mG. Therefore, the Zeeman shift is below 1~kHz. Finally, the second-order Doppler effect is estimated. Assuming the discharge temperature is between 300 and 400~K, the second-order Doppler shift is 7.2(10)~kHz. The overall experimental uncertainty is listed in Table~\ref{tab:table1}. Taking the frequency shift and uncertainties into account, the frequency of the 2$^{1}$S$_{0}$-3$^{1}$D$_{2}$ transition in $^{4}$He is determined to be 594~414~291~803(13)~kHz. This result is in good agreement with a previous determination, 594~414~291~935(334)~kHz, using the 2$^{1}$S$_{0}$-2$^{1}$P$_{1}$~\cite{luo2013} and the 2$^{1}$P$_{1}$-3$^{1}$D$_{2}$ transitions~\cite{luo2013a}, and the precision is improved by a factor of 25. We also investigate the linewidth of 2$^{1}$S$_{0}$-3$^{1}$D$_{2}$ transition shown in Fig.~\ref{fig:fw}. The natural linewidth of 2$^{1}$S$_{0}$-3$^{1}$D$_{2}$ transition is mainly contributed from the lifetime of the 3$^{1}$D$_{2}$ state. By extrapolating the linewidth to zero pressure and zero power, the measured linewidth is 11.21(12)~MHz which is in reasonable agreement with the square root of the quadratic sum of natural linewidth (10.36~MHz~\cite{wedding1990}) and  estimated transit time broadening linewidth (3.6(2)~MHz) using the atomic velocity and the waist beam size in the power built-up cavity. Combining the theoretical value of the 3$^{1}$D$_{2}$ ionization energy 365~917~749.02(2)~MHz from Ref.~\cite{drake2006} , we obtain the 2$^{1}$S$_{0}$ ionization energy: 960~332~040~866(24)~kHz. The contribution of the uncertainty is due to the theoretical calculation in 3$^{1}$D$_{2}$ state~(20~kHz) and experimental uncertainty~(13 kHz) listed in Table~\ref{tab:table1}. To compare with theories on the 2$^{1}$S$_{0}$ state, we subtract the nonrelativistic energy, the first relativistic corrections and the finite nuclear size correction calculated in Ref.~\cite{pachucki2017}, from the theoretical value of the 2$^{1}$S$_{0}$ ionization energy, 960~332~038.0(1.7)~MHz~\cite{pachucki2017}. The 2$^{1}$S$_{0}$ Lamb shift is deduced to be L(2$^{1}$S$_{0}$)~=~2806.864(24)~MHz. In Fig.~\ref{fig:comp}, we compare our determination of the 2$^{1}$S$_{0}$ Lamb shift with previous results by combining the theoretical values in 3D states, and the triplet-state measurements using the 2$^{3}$S$_{1}$-2$^{1}$S$_{0}$ measurement~\cite{van2011}. Our result is in good agreement with other determinations and is the most precise measurements at present, and it provides another independent test of the QED atomic theory. We can make another comparison to avoid the large theoretical uncertainty of 2S states. We combine our result with 2$^{3}$S$_{1}$-2$^{1}$S$_{0}$~\cite{van2011}, 2$^{3}$S$_{1}$-2$^{3}$P$_{0}$~\cite{pastor2012} and 2$^{3}$P$_{0}$-3$^{3}$D$_{1}$~\cite{luo2016} transitions to obtain the separation between 3$^{1}$D$_{2}$ and 3$^{3}$D$_{1}$ states, which has been of theoretical interest for the singlet-triplet mixing in 3D state. The result is in reasonable agreement with theoretical calculation~\cite{drake2006} and is listed in Table~\ref{tab:table2}. Furthermore, the Lamb shifts of the 2$^{1}$S$_{0}$ and 2$^{1}$P$_{1}$ states can be deduced by this work and previous measurement~\cite{luo2013}. All deduced Lamb shifts are listed in Table~\ref{tab:table3} along with the theoretical values.

\begin{table}[b]
\caption{\label{tab:table2}%
Comparison of the 2$^{1}$S$_{0}$-3$^{1}$D$_{2}$ and 3$^{3}$D$_{1}$-3$^{1}$D$_{2}$ intervals with theory (unit: MHz)}
\begin{ruledtabular}
\begin{tabular}{ldd}
 &\multicolumn{1}{c}{\textrm{2$^{1}$S$_{0}$-3$^{1}$D$_{2}$}}  &  \multicolumn{1}{c}{\textrm{3$^{3}$D$_{1}$-3$^{1}$D$_{2}$ }} \\
\hline
This work&594~414~291.803(13) &101 143 .889(29)\footnote{combined with Ref.~\cite{van2011,pastor2012,luo2016}}\\  
Theory&594~414~289. 8(1.9)\footnote{Ref.~\cite{drake2006,drake2008}}&101 143 .950(28)\footnote{Ref.~\cite{drake2006}}\\
Difference&2.9&0.61\\
\end{tabular}
\end{ruledtabular}
\end{table}

\begin{table}
\caption{\label{tab:table3}%
The Lamb shifts of 2$^{1}$S$_{0}$, 2$^{3}$S$_{1}$, and 2$^{1}$P$_{1}$ states (unit: MHz)}
\begin{ruledtabular}
\begin{tabular}{lddr}
 State&\multicolumn{1}{c}{\textrm{This work}}  &  \multicolumn{1}{c}{\textrm{Theory~\cite{pachucki2017}}} \\
\hline
2$^{1}$S$_{0}$&2806.864(25)\footnote{combined with Ref.~\cite{drake2006,pachucki2017}}&2809.7(1.7) \\  
2$^{3}$S$_{1}$ &4057.130(26) \footnote{combined with Ref.~\cite{drake2006,van2011,pachucki2017}}&4058.6(1.3)  \\
2$^{1}$P$_{1}$ &47.32(18) \footnote{combined with Ref.~\cite{drake2006,luo2013a,pachucki2017}}&48.87(40)  \\
\end{tabular}
\end{ruledtabular}
\end{table}

\section{\label{sec:level3} CONCLUSION}
In conclusion, the absolute frequency of the 2$^{1}$S$_{0}$-3$^{1}$D$_{2}$ two-photon transition of $^{4}$He at 1009~nm is measured for the first time and a precision of 13~kHz is achieved. Our result is more precise than previous determination using 2$^{1}$S$_{0}$-2$^{1}$P$_{1}$~\cite{luo2013} and 2$^{1}$P$_{1}$-3$^{1}$D$_{2}$~\cite{luo2013a} transitions by a factor of 25. A new determination of the ionization energy of the 2$^{1}$S$_{0}$ state is obtained by taking the theoretical value of the 3$^{1}$D$_{2}$ state. It is consistent with the best previous determination with 1.6 times improvement in precision. More importantly we can deduce the most precise Lamb shift of the 2$^{1}$S$_{0}$ state. In addition, the energy separation between the 3$^{3}$D$_{1}$ and 3$^{1}$D$_{2}$ states is deduced using present result and other previous measurements and it agrees with theoretical calculation. In the near future, the absolute frequency measurement for 2$^{1}$S$_{0}$-3$^{1}$D$_{2}$ two-photon transitions in  $^{3}$He will be performed to investigate the hyperfine structure of 3$^{1}$D$_{2}$ state in $^{3}$He and the isotope shift between $^{3}$He and $^{4}$He. The long-standing discrepancy of the $^{3}$He 3$^{3}$D$_{1}$-3$^{1}$D$_{2}$ separation~\cite{drake2006,derouard1980} will be resolved.

\begin{acknowledgments}
We thank Chunghwa Telecom Lab. for lending us a cesium clock. This project is supported by the Ministry of Science and Technology and the Ministry of  Education of Taiwan. L.-B. Wang acknowledges support from Kenda Foundation as a Golden-Jade fellow.
\end{acknowledgments}

\nocite{*}

\bibliography{apssamp}

\end{document}